\documentclass[11pt]{article}
\usepackage[utf8x]{inputenc}
\usepackage{amsmath} \usepackage{amsthm} \usepackage{amsfonts} \usepackage{amssymb}
\usepackage{array} 
\usepackage{booktabs}
\usepackage{enumitem}
\usepackage{indentfirst}
\usepackage{latexsym}
\usepackage{makeidx}
\usepackage{graphicx}
\usepackage{booktabs}
\usepackage[font=small,format=hang,labelfont=sf]{caption}
\usepackage{subfig}
\theoremstyle{plain} 
\usepackage[rightcaption]{sidecap}

\begin{document}

\title {Dynamical correlations in the escape strategy \\ of Influenza A virus} 
\author{Lorenzo Taggi$^{1,2}$, Francesca Colaiori$^{1,3}$, Vittorio
  Loreto$^{1,4}$ and Francesca Tria$^4$}

\maketitle

\centerline{$^1$Sapienza University of Rome, Physics Department, P.le
  A. Moro 5, 00185 Rome, Italy}

\centerline{$^2$ Max Planck Institute for Mathematics in the Sciences,
  Inselstr. 22, 04103 Leipzig, Germany}

\centerline{$^3$ CNR-ISC, P.le A. Moro 5, 00185 Rome, Italy}

\centerline{$^4$ ISI Foundation, Via Alassio 11/c, 10126, Torino,
  Italy}

\section*{Abstract}
\noindent 
The evolutionary dynamics of human Influenza A virus
  presents a challenging theoretical problem. An extremely high
  mutation rate allows the virus to escape, at each epidemic season,
  the host immune protection elicited by previous infections. At the
  same time, at each given epidemic season a single quasi-species,
  that is a set of closely related strains, is observed. A non-trivial
  relation between the genetic (i.e., at the sequence level) and the
  antigenic (i.e., related to the host immune response) distances can
  shed light into this puzzle. In this paper we introduce a model in
  which, in accordance with experimental observations, a simple
  interaction rule based on spatial correlations among point mutations
  dynamically defines an immunity space in the space of sequences.
  We investigate the static and dynamic structure of this space and we
  discuss how it affects the dynamics of the virus--host interaction.
  Interestingly we observe a staggered time structure in the virus
  evolution
  as in the real Influenza evolutionary dynamics. 

\section*{Introduction}
\noindent 
The interest of the scientific community in the Influenza A virus
evolution has been continuously increasing in the last
years~\cite{Ferguson_2003,Nelson_2007,Vijaykrishna_2011}.
Understanding the mechanisms driving the ever--changing of the
antigenic determinants is crucial in order to implement effective
prevention strategies. Major efforts have been devoted to explain
apparently contradictory features. On the one hand the virus mutates
fast enough so that the same host can be infected several times in the
course of its life, on the other hand a viral quasi-species can be
sufficiently well defined in any given epidemic season, so that a
temporarily effective vaccine can be developed. The peculiar
evolutionary dynamics of the Influenza A virus is revealed by the
comb--like shape of its phylogenetic
tree~\cite{Fitch_1997,Bush_1999,Bush_2000}, as reconstructed from
haemaglutinin (HA) coding sequences. It has been contrasted with
phylogenetic trees of other viruses~\cite{Grenfell_2003,Pompei_2012},
as measles virus and HIV virus at the population level, which show
more ramified patterns~\footnote{The phylogenetic tree of HIV virus
  inside a single host, where selective pressure plays a crucial role,
  presents instead features similar to those of the Influenza A virus
  tree.}.

A crucial mechanism driving the interaction between the virus and the
host immune system is \textit{cross--immunity}: after being infected
by a strain, the host acquires partial or total immunity to a set of
other strains {\it antigenically similar} to the infecting
one~\cite{Gill_1977}. However it is not yet clear what determines the
similarity relation in terms of genetic distance. A first attempt to
reproduce in a modelling framework the complex balance between strains
proliferation induced by antigenic drift, and strains selection,
induced by the increasing acquired immunity of the hosts, is due to
Ferguson et al.~\cite{Ferguson_2003}. In that work, a mechanism of
broad spectrum cross immunity, lasting for a period of several weeks
after infection, in addition to the life--long cross--immunity, is
claimed to be crucial in order to recover the observed evolutionary
dynamics of the Influenza A virus. Although this idea seems to be
confirmed in the framework of simple evolutionary
models~\cite{Tria_2005,Bianconi_2011}, a clear evidence of the
existence of such a mechanism has not been provided so far.

A common trait of the above mentioned and previous
models~\cite{Girvan_2002} is the assumed equivalence between genetic
and antigenic distance: mutations in the HA protein accumulate in time
until eventually the mutated strain becomes enough antigenically
distant to escape host immunity. In this case the degree of
cross--immunity between the two strains is measured in terms of the
Hamming distance between their sequences. Recent studies, however,
highlight how that assumption is not completely
correct~\cite{Smith_2004}: high genetic differences can be irrelevant
from the antigenic point of view and, vice versa, few nucleotidic
mutations can elicit a large antigenic
effect~\cite{Plotkin_2002,Smith_2004}, indicating that the
accumulation of genetic distance is not a necessary (and sometimes nor
sufficient) condition for the emergence of antigenically novel
strains. 

Further, it has been pointed out that amino acid changes which seem to 
be relevant in differentiating two specific antigenic clusters, can 
exhibit a null antigenic effect when appearing in different 
sequences~\cite{Smith_2004}, suggesting that antigenic clusters cannot 
simply be associated with key influential sites~\cite{Sanjuan_2004}.

The fact that antigenic distances could depend on the presence of
correlations among genetic mutations (epistasis)~\footnote{With the
  term epistasis we refer to the phenomenon through which the fitness
  effects of one mutation depend on the presence of other mutations in
  the genome.} might explain why phenotypic changes do not necessarily
appear as a consequence of accumulated mutations. Correlation between
mutations have indeed been observed~\cite{shih_2007} and the existence
of epistasis in neuraminidase (NA) and hemagglutinine (HA) proteins is
supported by phylogenetic and sequence
analysis~\cite{Plotkin2011,Bloom2010,Rimmelzwaan2005}. The effect on
the evolutionary dynamics of Influenza virus of a nontrivial relation
between genotypic and phenotypic (antigenic) space has been
investigated introducing the neutral network topology in the space of
sequences~\cite{newman_1998, Koelle_2006}. Neutral networks are
clusters of sequences connected by point-mutations which are
associated to the same phenotype, i.e., each sequence is antigenically
similar to the ones belonging to the same cluster and antigenically
different from the ones belonging to other clusters. In the model
proposed in \cite{Koelle_2006}, as a consequence of this specific
choice of genotypic-phenotypic mapping, Influenza evolution occurs by
``episodic selective sweeps''. When a mutated strain from a new
cluster appears, it has a small probability of being highly
advantageous. In that case, it fixes rapidly in the
population. Selective sweeps are thus triggered by rare events, and  followed by periods of neutral
evolution during which all the genomes observed in the population have
the same fitness.
However, recent genomic analysis of Influenza data presented
in~\cite{Strelkowa2012} supports a different evolutionary process of
Influenza, not compatible with the one described
in~\cite{Koelle_2006}. In this scenario Influenza evolution is driven
by a high supply of beneficial mutations that triggers ``clonal
interference''. Clones are sets of strains with similar sequences and
a common ancestor. Due to competition,
typically only lineages descending from a single high-fitness clone
will survive, while the others eventually will become extinct. The
expansion of a successful clone is driven by strongly beneficial
mutations which rapidly fix in the population. Such selective sweeps
reduce the diversity, though they never completely remove it, i.e.,
thanks to the high mutational rate, the population always remains multiclonal. 

In this paper we introduce a simple epistatic rule which defines a
genotypic-(antigenic) phenotypic mapping ``dynamically'' dependent on
spatial correlations among point mutations. 
Here correlations are ``dynamical'': neutral mutations at a certain point 
of the evolution are not established a priori as in~\cite{Koelle_2006},
rather they depend on all the past infections up to that moment. 
Carved 
out by our epistatic rule, one can then identify a cluster of
sequences respect to which the host is equally immune, that we call
epistatic immunity space. Thus, on the contrary of the mentioned above
neutral networks, our immunity space is not a static structure in the
space of sequences, rather it evolves dynamically self-consistent with
the virus-host interactions. This picture is compatible with the
results presented in~\cite{Strelkowa2012}, considering a high rate of
potentially beneficial mutations.
 We investigate the static and dynamical
properties of such an immunity space and then we point out how they
could affect the real virus dynamics. We first describe the non
trivial geometric and topological properties of the immunity space.
Then we consider a simple greedy dynamics that mimics the escape
strategies of a virus in an host population, relating in this way the
emerging structure of the immunity space to the viral evolutionary
dynamics. One striking consequence of the introduction of dynamically
correlated point mutations is the existence of a staggered time
structure in the virus evolution, characterised by an alternation of
periods where an high number of relatively low fitness strains are
able to spread the infection, followed by periods where a single
highly fit strain is the favoured escape mutant. The fitness is here
defined as proportional to the number of individuals not yet immune to
that strain. Interestingly, this behaviour is absent when the
antigenic distance is taken as directly proportional to the genetic
distance.

\section*{Modelling cross-immunity with an Epistatic Immunity Space} 
\noindent We
represent viral strains by binary sequences $\vec{v}$ of fixed length
$n$~\footnote{We identify the viral strain with its epitope sites by
  representing them consecutively in a unique connected sequence. The
  generalisation to a four letters alphabet will of course modify the
  quantitative results reported here, but should not affect our
  qualitative conclusions.}. We define the immunity set $I_n(\vec{v})$
of a strain $\vec{v}$ as the set of viruses antigenically similar to
it: those viruses that cannot infect a host that has been already
infected by $\vec{v}$. We can further consider the immunity elicited
by more than one strain, for instance by all the strains produced by
successive mutations and spread during an infection history. We call
the {\it Immunity Space}, $I_n(A)$, of the \textit{infection set} $A$
the union of all the immunity sets $I_{n}(\vec{v})$ of the strains in
$A$:
\begin{equation}
\label{math:EIspace}
I_n(A)= { \bigcup_{\vec{v} \in A} I_{n}(\vec{v})} .
\end{equation}
\noindent 
The immunity set depends on the definition of
\textit{antigenic similarity}. We here investigate the simplest choice
which includes correlations: we assume that two strains are
cross--immune unless they differ in at least two consecutive
bits. This choice is made for sake of simplicity, but any pair
of sites could be chosen without loss of generality and the present 
framework can be easily extended to more complex patterns of correlated mutations.
We thus consider from now on:
\begin{equation}
\label{math:EIset}
I_n(\vec{v}) \equiv \{\ \vec{z}\in H_n :\ z_{i}\neq v_{i} \Rightarrow
\ z_{|i+1|_n}= v_{|i+1|_n}\ \forall i \ \},
\end{equation}
where $H_n$ is the $n$--dimensional hypercube, composed of $2^n$
strings, with the metric given by the Hamming distance, and periodic
boundary conditions. We will call  $I_n(\vec{v})$ the {\it epistatic}
immunity set generated by the strain $\vec{v}$ and $I_n(A)$ the {\it Epistatic Immunity Space} (EIS) of the
\textit{infection set} $A$.

\section*{Static properties of the Epistatic Immunity Space} 
\begin{figure}[!htb]
\includegraphics[width=0.9\columnwidth]{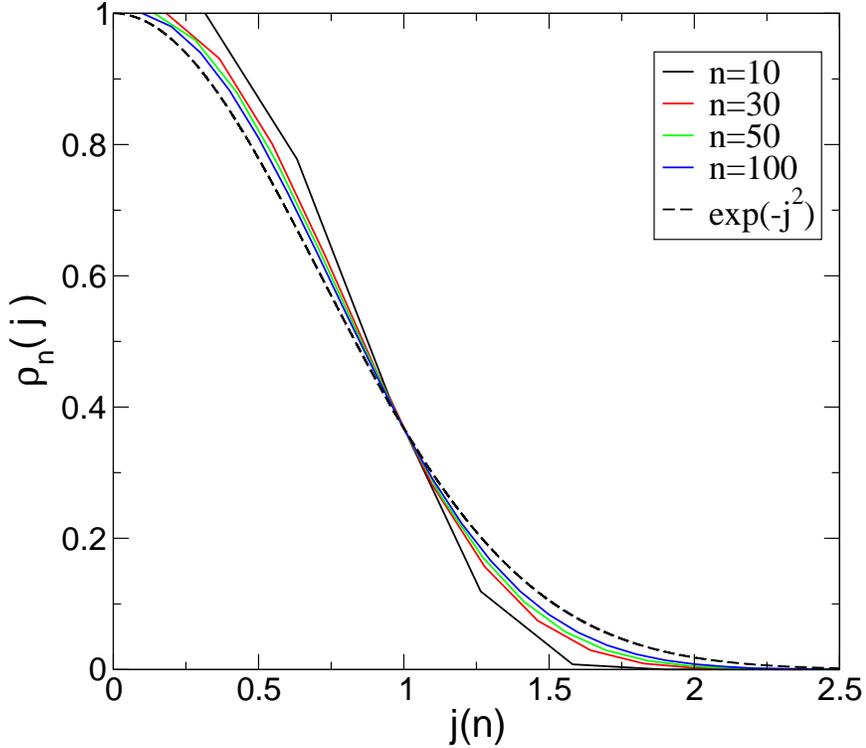}
\caption{Epistatic density function computed numerically for $n=10,
  30, 50, 100$. The densities are plotted as function of $j(n) =
  i/\sqrt{n}$, where $i$ is the Hamming distance. As $n$ increases,
  the epistatic density function converges to the well defined
  function $\rho_{\infty}(j) = exp(-j^{2})$.}
\label{fig:densepi}
\end{figure}

\begin{figure}[!htb]
    \includegraphics[width=0.9\columnwidth]{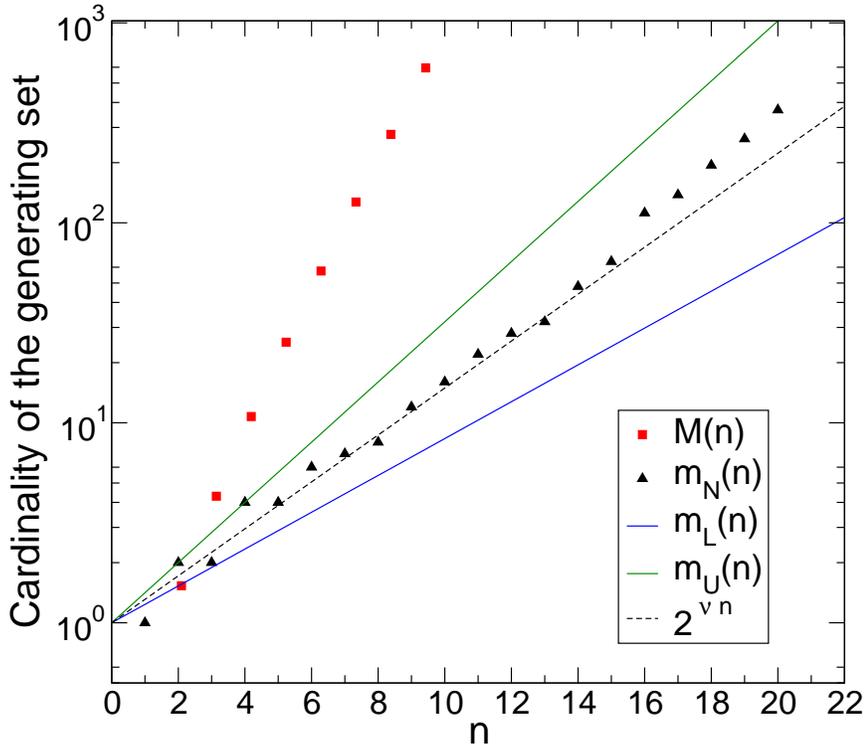}
    \caption{Numerical estimate of $m(n)$, $m_N(n)$, along with its lower and
      upper bound, and the value of $M(n)$ as a function of $n$. The
      dotted line represents the function $2^{\nu n }$, with $\nu =
      0.399 \pm 0.002$, which has been used to fit the first $15$
      values of $m_N(n)$.}
    \label{fig:stime}
\end{figure}

\noindent 
The fraction $\rho_{n}(i)$ of strains that belong to $I_n(\vec{v})$ and
have Hamming distance $i$ from $\vec{v}$ can be computed and reads
$\rho_{n}(i) \simeq \exp({-i^2 / n})$ (see Fig.~\ref{fig:densepi} for
the numerical plot and \cite{SI} for the analytical proof): on the one
hand correlations introduce a non trivial correspondence between
genotypic and phenotypic space, on the other hand antigenic similarity
is not completely uncorrelated from genetic
distance~\cite{Smith_2004}.
The size $S(n) \equiv|I_n(\vec{v})|$ of the immunity set generated by
a strain, i.e., the number of strains cross--immune to it, satisfies a
Fibonacci--like recursive relation: $S(n)= S(n-1) + S(n-2)$ with
initial condition $S(2)= 3$ and $S(3)= 4$. $S(n)$ is known as Lucas
sequence, and an explicit expression is known: $S(n) = \phi^n + (1 -
\phi)^n \simeq \phi^n$, where $\phi = (1+\sqrt{5})/2\sim 1.618$ is the
golden ratio, and the last asymptotic holds for large $n$ (see also
SI). The size $|I_n(A)|$ of the EIS generated by $k$ different strains
strongly depends on the actual form of the set $A$. Two quantities are
particularly relevant to provide bounds for every epidemic dynamics
with the above defined antigenic similarity measure: ({\it i}) $M(n)$,
the maximum number of distinct strings that fit in the sequence space,
and such that the next string would immunise the whole space (the
strings are therefore chosen with the maximum overlap between their
immunity sets); ({\it ii}) $m(n)$, the minimum number of strings
needed to immunise the whole sequence space, and therefore chosen with
the minimum overlap between their immunity sets. The computation of
$M(n)$ is straightforward: in order to have at least a string, say
$\vec{v}$, left out of the EIS, the infection set cannot contain {\it
  any} of the strings in $I(\vec{v})$. Therefore, the largest
infection set that does not immunise the whole hypercube is $A_{D(n)}
= H_n\, /\, I(\vec{v})$, which immunises the set $H_n / \{ \vec{v}
\}$, and $M(n) = 2^n - S(n)$. We estimate $m(n)$ by numerical
simulations and we provide analytically an upper $m_{U}(n)$ and a
lower $m_{L}(n)$ bound. A (trivial) lower bound is given by assuming
totally disjoint immunity sets, and it is given by counting the total
number of sequences divided by the size $S(n)$ of a single immunity
set: $m_{L}(n) \simeq 2^{n}/\phi^{n}=2^{\eta n}$,
with $\eta=1-\ln_2{\phi} \sim 0.306$. The fraction of strings
contained in the immunity set of a single strain is therefore
$2^{-\eta n}$. An upper bound $m_{U}(n)$ can be derived constructively
by exhibiting a set of sequences whose immunity sets cover the
sequence space. Such a set of sequences is obtained for example by
combining in all possible ways $n/2$ pairs of identical bits, either
$(0,0)$ or $(1,1)$ (for instance for $n=4$ such a coverage is realised
by the four sequences $( 0, 0, 0, 0), ( 1, 1, 0, 0), ( 0, 0, 1, 1), (
1, 1, 1, 1)$). The number of such sequences is $m_{U}(n)={2}^{\left
    [\frac{n}{2}\right ]}$, where $\left [\cdot\right ]$ denotes the
integer part. We estimate the asymptotic value of $m(n)$ numerically
by simulated annealing~\cite{Kirkpatrick_1983}. For any $n$ we look
for the set $A$ composed of $k$ sequences which minimises the cost
function $f_{n,k}(A)=2^n-|I_n(A)|$. $m(n)$ corresponds to the smallest
$k$ such that the minimum of the cost function $f_{n,k}(A)$ is equal
to $0$ (see \cite{SI} for a more detailed analysis). Our estimation is
$m(n)\simeq 2^{\nu n}$ with $\nu \simeq 0.4$, compatible with the
analytical bounds (see Fig.~\ref{fig:stime}).

Let us now focus on the topological properties of the EIS. Noticeably,
the EIS is always a connected set, for any infection history. To prove
this we need to show that for any pair of sequences ${\vec{x}}$,
${\vec{y}}$, there exists a path of cross-immune sequences joining
them. Since every single immunity set is connected, it is thus enough
to show that any pair of immunity sets overlap, or are at most
contiguous. Take $\vec{x}=\vec{0}$ without loss of generality, and
$\vec{y}=(y_1,y_2,\dots,y_n)$. For $n$ even, the two immune sets
always overlap at least in the sequence $(0,y_2,0,y_4,0,\dots,0,y_n)$.
For $n$ odd they are contiguous in the two points
$(0,y_2,0,y_4,0,\dots,0,y_{n-1},0)\in I_n(\vec{0})$ and
$(0,y_2,0,y_4,0,\dots,0,y_{n-1},y_n)\in I_n(\vec{y})$ (actually they
always overlap at some point unless $\vec{y}=\vec{1}$).

\begin{figure}[htp]
\includegraphics[width=0.9\columnwidth]{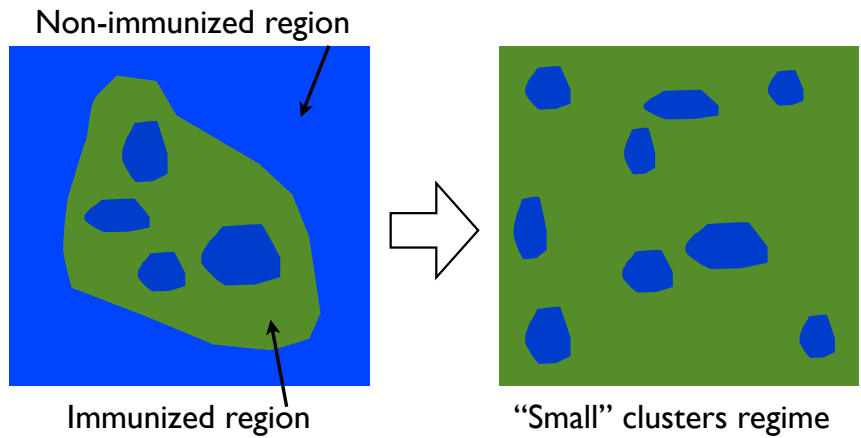}
 \caption{Sketch of the noninfectious (green) and infectious (blue)
   region of the sequence space. {Left}: for small $k$ above threshold
   the infectious region features a large connected cluster,
   corresponding to a infectious region of the hypercube, along as
   many small connected clusters. {Right}: increasing $k$ only small
   holes in the EIS are left.}
 \label{fig:cartoon}
\end{figure}

Though always connected, the EIS is not always simply connected, and
the complementary set, i.e., the infectious region, can be not
connected. This might have a strong impact on the underlying
virus--host interaction. For example, when $k$ strings are drawn at
random, the infectious region can be broken down in clusters only if
$k \geq\lceil n/2\rceil $ (see SI). For $k$ slightly above this
threshold the infectious region is composed by one big connected
cluster and many small connected clusters (``holes'' in the EIS). The
disappearance of the big connected cluster as $k$ increases sets the
threshold where a further spread of an epidemic is inhibited (see
Fig.~\ref{fig:cartoon} for a cartoon and \cite{SI} for a more detailed
analysis).

\section*{Simple dynamics on the Epistatic Immunity Space} 
\noindent 
We have so far
examined general topological properties of the EIS from a static point
of view. The nontrivial shape of the set of cross--immune sequences
can be however better highlighted considering simple infectious
dynamics. We then consider a local maximisation (LM) of the EIS: 
starting with a random strain, we choose at every 
step the next strain among those not already belonging to the EIS, and 
such that it maximises the size of the current EIS (and thus minimises the overlap
with the existing EIS). In case that several strings satisfy this
criterion we choose one at random among them. We iterate until the
whole space $H_n$ is noninfectious. 
Each step of the LM process corresponds to a new infection of the same
host population from the virus. Although virus evolution happens by
point mutations, when a virus infects again the same host it presents
multiple mutations with respect to the genome of the previous
infection. In fact the largest part of the population is typically
infected once every one or more epidemic seasons, while the point
mutation rate of the virus is much higher. This local maximisation
dynamics represents an attempt to model an effective interaction
between a population of viruses and a population of hosts who is more
likely to get infected provided the mutated virus is more
antigenically dissimilar from the previous successful one. From this
perspective this LM process mimics a successful escape strategy of
the virus in a host population in a coarse grained way in order to to
capture the implications of the adoption of the epistatic rule.

\begin{figure}[thtp]
\includegraphics[width=\columnwidth]{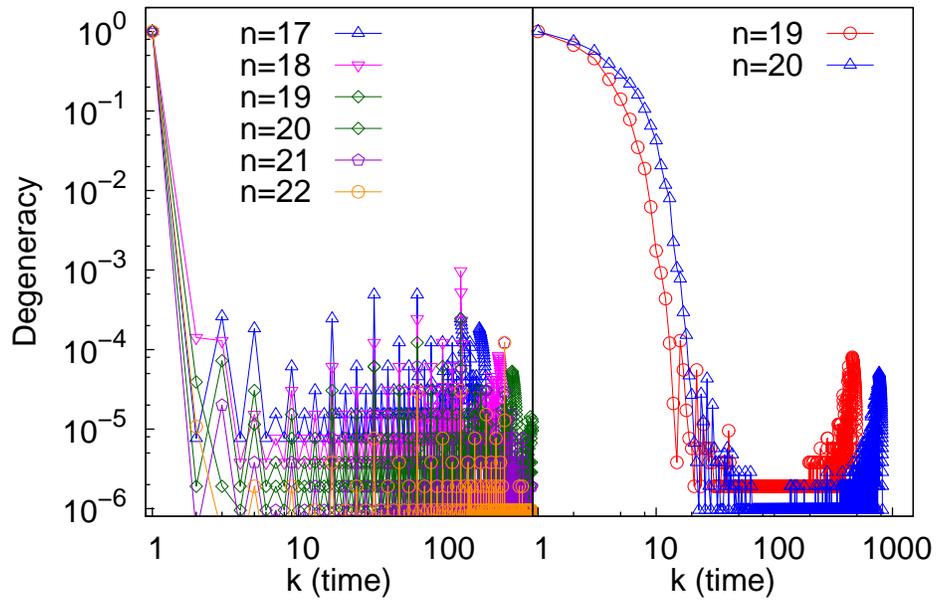}
\caption{{Left:} Degeneracy of strings allowed by the local
  optimisation dynamics (normalised with $2^n$) as a function of the
  iteration number $k$ (time) for the epistatic rule. The averages are
  taken over $1000$ realisations. {Right:} For comparison: same
  dynamics, but with cross--immunity defined by the Hamming rule with
  distance $D=4$ (the immunity set is the set of all strings whose
  Hamming distance $\leq D$ from the generating one).}
\label{fig:degeneracy}
\end{figure}

If we look at the number of sequences that satisfy the local
maximisation constraint at each time step, we find a peculiar
behaviour that is not observed when the immunity sets are constructed
by means of the bare Hamming distance from the generating strain. The
time behaviour features a well defined series of peaks corresponding
to an alternation of periods with many equivalent options (i.e.,
possible strategies for the virus) and only one optimal option to
maximise the immunity set (Fig.~\ref{fig:degeneracy}). This gives a
hint of how dynamical constraints arise from the presence of epistatic
interactions with respect to the case in which antigenic distance is
directly proportional to genetic distance.

\begin{figure}[htp]
\includegraphics[width=\columnwidth]{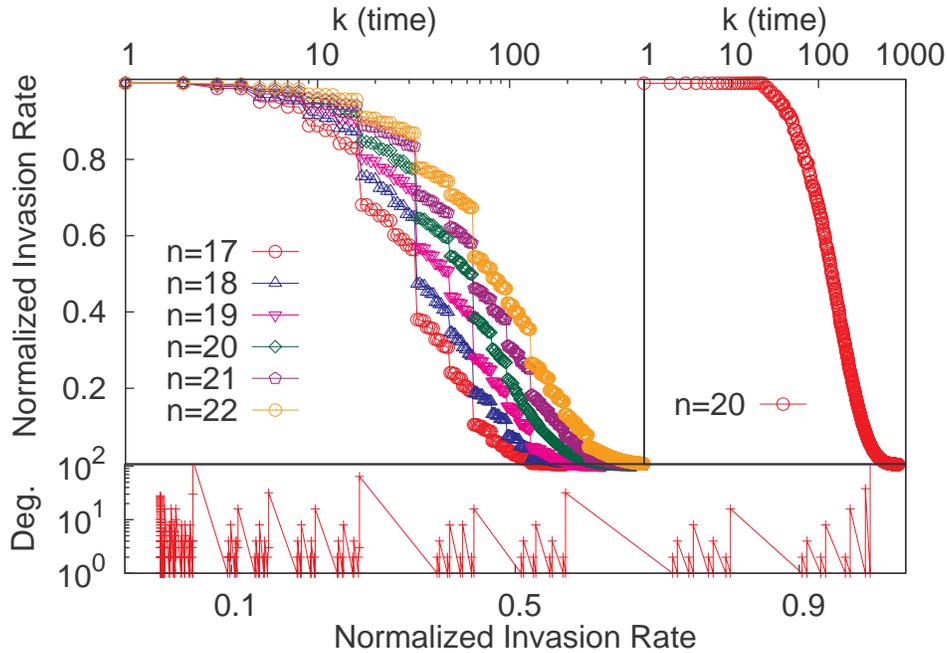}
  \caption{ {Top left:} Time behaviour of the normalised invasion rate,
    i.e., the fraction of sequences becoming noninfectious at time $k$
    for different values of $n$. {Top right:} For comparison we report
    the same quantity as in the Top left panel, but with
    cross--immunity defined by the Hamming rule. In this case no jumps
    are observed. {Bottom:} Parametric plot of the degeneracy vs. the
    normalised invasion rate for the epistatic rule with $n=19$.}
  \label{fig:salti_epi}
\end{figure}

To further characterise the epidemic dynamics we look at the {\it
  normalised invasion rate}, i.e., the fraction of strains becoming
noninfectious at each step of the LM dynamics
(Fig.~\ref{fig:salti_epi}). This quantity also shows a non-trivial
behaviour characterised by a series of hierarchically distributed
jumps that occur always at the same time steps, independently of $n$,
and that are not present when the same dynamics is studied with a
Hamming rule for cross--immunity (see Fig.~\ref{fig:salti_epi},
right). This points again to a staggered time structure with an
alternation of periods of highly effective immunisation, followed by
periods with a relatively lower immunisation rate. This picture is
also confirmed by the parametric plot in the bottom of
Fig.~\ref{fig:salti_epi} where the degeneracy (the fraction of optimal
strains) is plotted versus the normalised invasion rate. The peculiar
triangular structure, absent in the Hamming case, is the signature of
an alternation of times with no degeneracy (only one option)
corresponding to a high invasion rate followed by times with a very
high degeneracy and low invasion rate. This behaviour can be related
to the comb-like shape of the Influenza A phylogenetic tree, where a
single quasi-species is responsible for each annual epidemic and
antigenic clusters follow one another each few
years~\cite{Plotkin_2002}.

\section*{Conclusion and perspectives}
\noindent 
In this paper we focused on the
long-standing puzzle behind the strategies of viruses trying to escape
the immune system. We introduced in particular a model in which
cross--immunity, i.e., the mechanism through which a host acquires
partial or total immunity to a set of other strains {\it antigenically
  similar} to the infecting one, is defined in terms of dynamically
correlated point mutations. We have investigated how this {\em
  epistatic} rule carves in a non-trivial way the immunity space of
the host, i.e., the set of viruses to which a host is immune after
infection by all the strains in his infection history. We quantified
the geometrical and topological properties of this space, highlighting
qualitative differences with respect to the case when a distance that
disregards correlations among different sites (e.g., the Hamming
distance) is considered. We have further studied a simple greedy virus
dynamics, focusing on the important differences with respect to the
case where the usual Hamming distance defines cross--immunity. Here we
obtained the striking result that a simple escape virus dynamics on
the {\em epistatically} carved immunity space, leads to a staggered
time structure of the virus evolution. Times where one single choice
exists that maximises the invasion rate are followed by times where
many different options exist to immunise a relatively smaller set of
sequences. This results contrasts with the corresponding result
obtained without dynamical correlations in the definition of the
cross-immunity. Although obtained in the framework of a toy model, it
is quite tempting to identify our staggered time structure with the
succession in time of different antigenic clusters and with the more
violent epidemic outbreaks at each cluster change, as observed in real
virus-host dynamics. The analysis presented here can help
understanding the effect of the conjectured epistatic interactions on
the shape of immunity clusters as well as on the viral evolutionary
dynamics at large. This in turn can trigger the investigation of more
realistic virus-host interaction schemes incorporating the epistatic
rule.



\end{document}